\documentclass[twocolumn,showpacs,preprintnumbers,amsmath,amssymb]{revtex4}

\usepackage{graphicx}
\usepackage{dcolumn}
\usepackage{bm}
\usepackage{epsfig}

\begin{document}

\preprint{Physical Review Letters (2007) Vol. 98, 238104}
\title{The effect of salt concentration on the electrophoretic speed of a polyelectrolyte through a nanopore}

\author{Sandip Ghosal}
\affiliation{%
Northwestern University, Department of Mechanical Engineering\\
2145 Sheridan Road, Evanston, IL 60208
}%


\date{\today}

\begin{abstract}
In a previous paper~\cite{ghosal_PRE06} a hydrodynamic model for determining the electrophoretic speed of a polyelectrolyte
through an axially symmetric slowly varying nanopore was presented in the limit of a vanishingly small Debye length. 
Here the case of a finite Debye layer thickness is considered while restricting the pore geometry to that of 
a cylinder of length much larger than the diameter. Further, the possibility of  a uniform surface charge on the walls
 of the nanopore is taken into account.
It is thereby shown that for  fixed $\zeta$-potentials on the surface of the polyelectrolyte and on the pore wall, 
the electrophoretic speed is independent of the Debye length. The translocation speed depends on the salt 
concentration only to the extent that the $\zeta$-potentials depend on it, and further, this dependence is very weak. 
It is shown that the calculated transit times are consistent with recent measurements in silicon nanopores 
that reveal this insensitivity to salt concentration. 
 \end{abstract}

\maketitle

The translocation of polymers across nanometer scale apertures in cell membranes is a common phenomenon in 
biological systems~\cite{Alberts}. If the polymer carries a charge, an applied electric potential can drive the translocation.
The change in electrical conductance of a single nanopore as a polymer transits the pore can be reliably detected 
and used to characterize the polymer~\cite{kasianowicz_PNAS96}.
 A number of experimental studies~\cite{meller_etal_PNAS00,ssdna_sequence_nbt,storm_nature,storm_physRevE05}
  as well as a few theoretical ones~\cite{lubensky_nelson,ghosal_PRE06}  on the electrically driven 
  translocation of polymers across nanopores 
have appeared recently. Interest in the phenomenon is to a large extent motivated by the possibility of refining it to the point where 
the base sequence of a DNA strand can be read with single base resolution 
as the DNA transits the pore~\cite{deamer_trendsinbiotech}. This would provide a sequencing method that is faster and 
cheaper than existing ones by many orders of magnitude. A technological challenge is the trade off between noise and resolution.
In typical experiments with solid state nanopores a single base pair transits the pore in about $\sim 10^{-8}$ sec -- much too short to be 
resolved. On the other hand the voltage across the pore cannot be sufficiently reduced to slow down the DNA because then 
the change in current would not be detectable above the noise. A theoretical analysis of the problem to determine how the translocation 
speed depends on the controllable parameters is  therefore of value in guiding the experimental work. 

In an earlier paper~\cite{ghosal_PRE06} (henceforth Paper A) a hydrodynamic model  was proposed for describing the 
process of electrically driven translocation across the nanopore. The speed of translocation
is determined by a balance of electrical and viscous forces 
arising from within the pore with proper accounting for the co- and counter-ions in the electrolyte.
The underlying  physics is not unlike  that of electrophoresis 
of small charged particles in an applied electric field except that here the proximity of the 
pore walls play an important role. The translocation speed was explicitly calculated for 
cylindrically symmetric pores by 
assuming an infinitely thin Debye layer and slowly varying pore 
radius. The calculated translocation speed 
was shown to be in close agreement with experimental measurements~\cite{storm_physRevE05} in solid state 
nanopores. The assumption of infinitely thin Debye layers was justified because of the high 
concentration of salt (1 M  KCl) in the electrolyte used in the experimental work. 
 More recently Smeets {\it et al.}~\cite{dekker_nano_lett06} have published experimental data 
 on a solid state nanopore  for an electrolyte 
with KCl concentration varying from 50 mM to 1.0M. Remarkably,
it was found that the most probable translocation time either did not vary at all 
with salt concentration or the variation was too small to be detected. 
In this paper the translocation speed is calculated 
based on the mechanism proposed in Paper A but allowing 
for a finite Debye Layer thickness while restricting the geometry to a long cylindrical pore. 
 The objective is to determine whether the proposed hydrodynamic  model is consistent with the observed 
experimental dependance of the translocation speed on salt concentration.

\begin{figure}
    \includegraphics[angle=0,width=3.0 in]{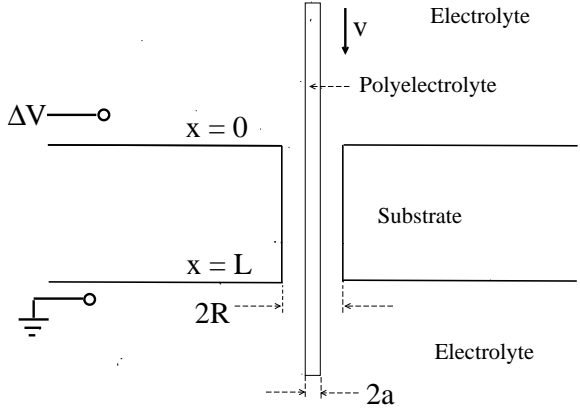}
    \caption{Geometry of the pore region.}
    \label{geom_nanopore2}
\end{figure}
Figure~\ref{geom_nanopore2} shows the geometry for our simplified calculation. The 
pore shape in the experiment actually resembles a hyperboloid with a  smallest diameter of $10.2$ nm. 
Smeets {\it et al.}~\cite{dekker_nano_lett06}
report that the bulk conductance of the pore is equivalent to that of a cylindrical 
nanopore of identical diameter and length $L=34$ nm. For the purpose of comparing 
our calculation with experiments, we will consider a cylindrical pore with these dimensions. 
Moreover, we will assume the flow field to be uniform in the axial direction, an assumption 
that is strictly valid only for an infinitely long cylinder. 
Let us model the part of the  polyelectrolyte inside the pore by a straight rigid cylindrical rod (of radius $a=1$ nm) that 
is co-axial with the cylindrical pore (of radius $R=5.1$ nm) and translocating at a velocity $v$ in the axial 
direction ($x$). Such a model is reasonable since the persistence length of double stranded DNA (ds-DNA) is about $50$ nm.
In the absence of an applied pressure gradient (the inlet and outlet reservoirs are both at atmospheric pressure), the flow velocity 
$\mathbf{u} = \hat{\mathbf{x}} \, u(r)$ in the region between the polyelectrolyte and the pore 
wall satisfies Stokes equation (the Reynolds number $\text{Re} \sim 10^{-4}$): 
\begin{equation} 
\mu \nabla^{2} \mathbf{u}  + \hat{\mathbf{x}} \rho_{e} (r)\,  \, E_{0} = 0 
\label{eq:stokes}
\end{equation} 
where $\mu$ is the dynamic viscosity of the electrolyte, $E_{0}$ is the applied electric field 
in the pore and $\rho_{e}(r)$ is the electric charge distribution 
due to the co and counter-ions as a function of the distance from the axis ($r$).
 On account of axial symmetry of the cylindrical geometry 
considered, the movement of ions due to the current or  bulk motion does not change the electric charge density 
in the diffuse layer. Thus, $\rho_{e} (r)$ is both the equilibrium charge density ($E_{0}= v =0$) as 
well as the charge density after a potential difference is applied across the cylinder ($E_{0}$ and $v$ $\neq 0$).
Poisson's law relates $\rho_{e}$ to the electric potential $\phi$:
\begin{equation} 
\epsilon \nabla^{2} \phi = - \rho_{e} 
\label{eq:poisson}
\end{equation} 
where $\epsilon$ is the dielectric constant of the electrolyte. Equations (\ref{eq:stokes}) and 
(\ref{eq:poisson}) must be solved with boundary conditions 
\begin{eqnarray} 
\phi (r=a) &= & \zeta_{p} \label{bc:phi_a}\\
\phi (r=R) &=& \zeta_{w} \label{bc:phi_R}
\end{eqnarray} 
where  $\zeta_{p}$ and $\zeta_{w}$ are the $\zeta$-potentials 
at the surface of the polyelectrolyte and the wall respectively. The classical condition of no-slip at the walls 
imply that 
\begin{eqnarray} 
u(r=a) &=& v \label{bc:u_a}\\
u(r=R) &=& 0.       \label{bc:u_R}
\end{eqnarray} 
Equations (\ref{eq:stokes}) and (\ref{eq:poisson}) imply that the function 
$f = u - \epsilon E_{0} \phi / \mu$ satisfies the equation 
\begin{equation} 
\nabla^{2} f = \frac{1}{r} \frac{d}{dr} \left( r \frac{df}{dr} \right) = 0 
\end{equation} 
with boundary conditions 
\begin{eqnarray} 
f(a) = v - \frac{\epsilon E_{0} \zeta_{p}}{\mu} \\
f(R) =  - \frac{\epsilon E_{0} \zeta_{w}}{\mu}.
\end{eqnarray} 
The equation for $f$ is readily integrated, giving us the flow profile $u(r)$:
\begin{equation} 
\frac{u(r)}{u_{e}} =  \frac{\phi - \zeta_{w}}{\zeta_{p}} + \left( \frac{v}{u_{e}} + 
 \frac{\zeta_{w}-\zeta_{p}}{\zeta_{p}} \right) \frac{ \ln (r/R) }{\ln (a/R) } \label{velocity}
\end{equation}
where $u_{e} = \epsilon E_{0} \zeta_{p} / \mu$ is a characteristic electrophoretic 
velocity. The potential $\phi$ is determined from the Poisson-Boltzmann equation 
which is obtained on substituting the Boltzmann distribution on the right hand 
side of equation (\ref{eq:poisson}): 
\begin{equation} 
\epsilon \nabla^{2} \phi = - \sum_{k}  e z_{k} n_{k}^{(\infty)}  \exp{\left( - \frac{z_{k} e \phi}{k_{B} T} \right) }.
\end{equation} 
Here $z_{k}$ is the valence and  $n_{k}^{\infty}$  the far field concentration 
of ion species $k$, $e$ is the magnitude of the electronic charge, $k_{B}$ is the Boltzmann 
constant and $T$ the absolute temperature of the electrolyte.
However, for the purpose of determining the  translocation speed $v$, an explicit 
solution for $\phi$ will not be needed. The required velocity is obtained from the 
condition that the total force on the section of the polymer inside the pore is zero: 
\begin{equation}
F_{e} + F_{v} = 0 
\label{eq:equilibrium}
\end{equation} 
where $F_{e}$ and $F_{v}$ are the electric and viscous forces per unit length 
of the polymer. If $\lambda$ is the charge per unit length of the polymer then 
\begin{equation} 
F_{e} = \lambda E_{0} = - 2 \pi a \epsilon E_{0} \phi^{\prime}(a)
\label{eq:F_elec}
\end{equation} 
by Gauss's law. 
The viscous force, $F_{v} = 2 \pi a \mu u^{\prime}(a)$  can be calculated using (\ref{velocity}): 
\begin{equation} 
\frac{F_{v}}{ 2 \pi \mu u_{e}}
=  a \frac{\phi^{\prime}(a)}{\zeta_{p}} 
  + \left( \frac{v}{u_{e}} + \frac{\zeta_{w} - \zeta_{p} }{ \zeta_{p}} \right)
  \frac{1}{\ln{(a/R)} }.
  \label{eq:F_visc}
\end{equation} 
On substituting equations (\ref{eq:F_elec}) and (\ref{eq:F_visc}) into (\ref{eq:equilibrium}) we get 
\begin{equation} 
v = u_{e} \left( 1 - \frac{\zeta_{w}}{\zeta_{p}} \right) = \frac{\epsilon E_{0} \zeta_{p}}{\mu}
- \frac{\epsilon E_{0} \zeta_{w}}{\mu}.  \label{eq:v}
\end{equation}
Equation (\ref{eq:v}) for the translocation speed is the main result of this paper.
 Surprisingly, equation (\ref{eq:v}) is identical to the result we would have obtained 
if we had made the assumption of infinitely thin Debye layers as we did in Paper A. 
To see this, observe that  equation~(\ref{eq:v}) together with the Helmholtz-Smoluchowski slip 
boundary condition would imply a uniform flow $u(r) = - \epsilon \zeta_{w} E_{0} / \mu$
in the fluid which would satisfy  the condition of zero force on the polyelectrolyte considered together 
with its Debye layer.  Equation (\ref{eq:v}) also follows directly (with $\zeta_{w}=0$) from equation (12)
for the translocation speed in Paper A if one assumes a uniform cylinder for the pore shape. 
In addition, it has the following very simple interpretation in the limit of thin Debye layers: 
the part $-  \epsilon E_{0} \zeta_{w} / \mu$ is the electro-osmotic flow through 
the pore generated by the applied field and $\epsilon E_{0} \zeta_{p} / \mu$ is simply 
the electrophoretic speed of an object of arbitrary shape in  a reference frame fixed to the 
moving fluid in the nanopore~\cite{Russel}. 

\begin{figure}
    \includegraphics[angle=0,width=3.0 in]{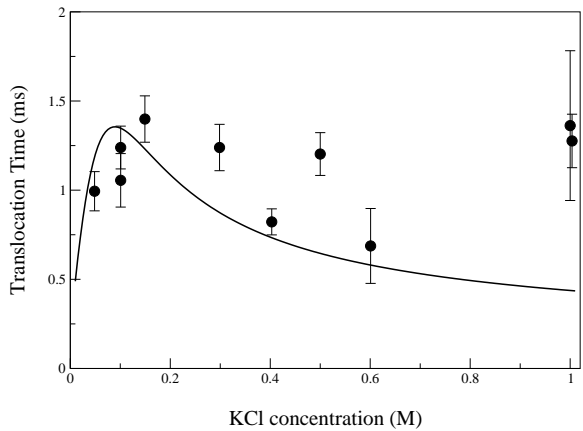}
    \caption{Translocation times for  $16.5$ $\mu$m long ds-DNA through a $10.2$ nm diameter solid state 
    nanopore. Solid line is calculated from equations (\ref{eq:v}), (\ref{zetap}) and (\ref{zetaw}), the symbols are 
    replotted from the data presented in Figure~4(b) (inset) of Smeets {\it et al.}~\cite{dekker_nano_lett06}}
    \label{time_nanopore2}
\end{figure}
Equation (\ref{eq:v}) will now be compared to experimental data due to 
Smeets {\it et al.} referenced earlier~\cite{dekker_nano_lett06}. In the Debye-Huckel approximation 
 the $\zeta$ potential of the polyelectrolyte, $\zeta_{p}$ is related to its 
 linear charge density $\lambda$ through the formula (see Paper A)
\begin{equation} 
\zeta_{p}  = \frac{\lambda \lambda_{D} }{2 \pi a \epsilon} \,
\frac{K_0 (a / \lambda_D )}{ K_1 (a / \lambda_D ) },
\label{zetap}
\end{equation} 
where $\lambda_{D}$ is the Debye length and $K_{n}$ are the modified Bessel functions of order $n$. 
For a univalent salt like KCl the Debye length (in nm) is given by~\cite{Probstein}
$\lambda_{D} = 0.303 / \sqrt{c}$ where $c$ is the Molar concentration of the salt. The experimental 
data is in the range $0.05$ to $1.0$ M so that $\lambda_{D}$ ranges from $0.30$ nm to $1.36$ nm. 
Since $R=5.1$ nm and $a=1.0$ nm, there is no significant overlap between the 
 Debye layers at the polyelectrolyte and the wall for concentrations above $0.05$ M, 
 though for even smaller concentrations such effects may be expected.
 Thus, it is reasonable to use the expression (\ref{zetap}) 
 which is strictly true only for an isolated infinite rigid rod in an unbounded electrolyte. 
 The dielectric constant $\epsilon / \epsilon_{0} = 80$ and dynamic viscosity 
$\mu = 8.91 \times 10^{-4}$ Pa s for the electrolyte are taken as those of water.
For the linear charge density on the DNA we take  $5.9$ electronic charges per nm reduced by 
the Manning factor of $4.2$, thus $\lambda = -2.25 \times 10^{-10}$ C/m. This assumption 
is supported by recent force measurement experiments \cite{direct_force_DNA06} that show 
that polyelectrolyte charge  is reduced by the classical Manning factor when the DNA is inside 
the pore over a wide range of salt concentrations. The electric field intensity is obtained 
by assuming that the entire voltage drop of $120$ mV occurs over the length of the 
equivalent cylinder which is $L=34$ nm, thus, $E_{0}=-3.53 \times 10^{6}$ V/m. 
The $\zeta$-potential at the $\text{SiO}_{\text{2}}$ wall may be obtained from the 
expression 
\begin{equation} 
\zeta_{w} = a_{0} - a_{1} \log_{10}{c}
\label{zetaw}
\end{equation}
where $c$ is the molar concentration of $K^{+}$ ions. The functional form of the dependence on concentration follows 
in the low counter-ion concentration limit from the nonlinear Gouy-Chapman model of the Debye layer 
in case of symmetric electrolytes. However, it 
has been shown to provide a good empirical fit to 
experimental data for counter-ion concentrations up to 1.0M~\cite{zetareview_eph04a}.
 For KCl on silica  $a_{0} \approx 0$ and $a_{1}  \approx - 30$ mV. 
 
 The translocation velocity, $v$ is  calculated from equation~(\ref{eq:v}) 
for a range of concentrations from $0.01$ to $1.01$M. The corresponding translocation 
time for a $L_{p}=16.5$ $\mu m$ long DNA, $t = L_{p}/v$ is shown as the solid line in 
Figure~\ref{time_nanopore2}. A notable feature is the lack of sensitivity of the translocation time  to 
the salt concentration: it changes by at most a factor of three when the salt concentration 
ranges over two orders of magnitude. 
Taking into account the considerable scatter in the experimental data and the 
various approximations made in the theory, the agreement between the two is quite reasonable,
pointing to the adequacy of the underlying  hydrodynamic model. The existence of a 
maximum in the translocation time at a concentration of about $0.1$ M KCl seems to be 
supported by the data, although one cannot be completely certain of this on account of the uncertainty in the data. 
The principal uncertainties involved in applying the hydrodynamic model to nanopores were discussed 
in Paper A. Those same considerations apply to the current calculations as well and need not be repeated here. 
It should also be kept in mind that although the motion of the polymer is treated as a unidirectional  translation 
at constant speed, the actual translocation takes place via a drift diffusion process as described
by Lubensky and Nelson~\cite{lubensky_nelson}. 
Here it is assumed, as is done in the classical theory of Brownian motion of particles, 
 that, the mean part of the motion of the polymer may be obtained through the solution of a classical hydrodynamics 
 problem that ignores the fluctuating forces. 
The hydrodynamic model or indeed any model that localizes the entire resistive force at the pore region would predict a translocation 
speed that is independent of polymer length. This is valid only for polymers that are not too long (see Paper A). For very long polymers 
the resistive force has an entropic part as discussed by various authors~\cite{sung_park,Boehm99,muthukumar99}.
St\"{o}rm {\it et al.}~\cite{storm_nanolett05} have suggested that the viscous drag on the randomly coiled part 
of the polymer lying outside the pore could also be significant.

DNA translocation experiments that have been performed to date can be divided into two classes; those that use a
natural protein nanopore ($\alpha$-hemolysin) on a lipid membrane~\cite{kasianowicz_PNAS96,meller_etal_PNAS00},
 and those that use a mechanical nanopore on a solid substrate made by specialized techniques~\cite{storm_nature,lietal_nature01}.
 Although the principle is similar, these two types of nanopores differ with respect to some important 
 details. One essential physical difference is that the narrowest part of the $\alpha$-hemolysin pore is about 
 $1.5$-$2.0$ nm in diameter so that only single stranded DNA or RNA is able to pass through it. When the pore is blockaded 
 by such a single strand the blockade is almost complete in that very few ions and probably none of the water 
 is able to pass through the blocked pore. Although solid state pores can be made with pore sizes approaching 1 nm, most 
 of the experiments to date have been done with $5-10$ nm diameter pores which can be made in a more 
 reliable and reproducible manner. These larger diameter pores admit both single and double stranded DNA, 
 and furthermore dsDNA can enter the pore in a folded fashion, notwithstanding the relative rigidity of these 
 polymers~\cite{storm_physRevE05}. The main observable difference in terms of translocations across the 
 two kinds of pores is that the polymer passes through the solid state pores about two orders of magnitude faster 
 than it does through $\alpha$-hemolysin pores. 
  It is important to stress that the analysis presented here applies to only the $5-10$ nm solid state 
 nanopores. Although a similar hydrodynamic model could be constructed to model the viscous force arising 
 out of the water in the vestibular part of the $\alpha$-hemolysin pore, such a model must of necessity differ from the current 
 one in the details of its formulation.  Furthermore, the applicability of the continuum equations for electrostatics and hydrodynamics would be 
 questionable to a much greater degree than in the analysis presented in this paper. It has been suggested that 
 in order to explain the much slower translocation speed in protein pores, something other than hydrodynamics is needed:
 perhaps an atomic level pore-polymer interaction, an electrostatic self-energy barrier~\cite{meller_etal_PRL06}
 or the energy cost associated with stripping hydration layers from the polymer as it enters the pore. 
  The results derived in this paper neither supports nor refutes the validity of these alternate mechanisms for the $1.5-2.0$ nm protein pores. 
 It does however show that for the $5-10$ nm  solid state pores hydrodynamic resistance can explain the 
 experimental data in the absence of any of the other mechanisms.

In conclusion, the hydrodynamic model introduced in Paper A to calculate the average transition time of a polyelectrolyte 
across a nanopore under an applied electric field was extended to treat the case of a finite Debye layer thickness, though 
the geometry was restricted to the simple case of a cylindrical pore. The predicted translocation times are found to be consistent
with available experimental data  to within the uncertainties inherent in the experiment and the theory. As a final remark, 
it is worth noting a few practical implications of the simple model presented here in relation to the problem of how one needs 
to tune the available parameters to make the translocation time as large as possible. First, 
Figure~\ref{time_nanopore2} shows that an optimal salt concentration exists  for which the translocation speed is a 
maximum, though the gain here is no more than a factor of $3$. 
 A better strategy is suggested by equation (\ref{eq:v}) which shows that $v$ vanishes 
if $\zeta_{w} = \zeta_{p}$. Physically this essentially amounts to balancing the electrophoretic migration of the DNA 
against an opposing electroosmotic flow generated at the wall. In principle this could be achieved by using an alternate 
substrate, a coating on the existing substrate or a physical or chemical treatment of it that alters its $\zeta$-potential. 
The object is to select a substrate such that $\zeta_{w} \approx \zeta_{p}$ and then ``fine tune'' the salt concentration to 
achieve a closer match. As an example, Poly(methyl methacrylate) (PMMA) is a commonly used substrate in microfluidic application for 
which $a_{0} = -4.06$ mV and $a_{1}=-12.57$ mV~\cite{zetareview_eph04b}. Using these values in (\ref{zetaw}) and plotting the result together 
with equation (\ref{zetap}) it is easily seen that the two curves intersect at a salt concentration of about $0.6$ M. 
Operating near this molarity with a PMMA substrate should result in significantly slower translocations. 

\bibliographystyle{/Users/sgh219/Work/LIBRARY/BIBFILES/prsty}
\bibliography{/Users/sgh219/Work/LIBRARY/BIBFILES/membrane,/Users/sgh219/Work/LIBRARY/BIBFILES/microfluidics}
\end{document}